\documentclass[%
    superscriptaddress,%
    twocolumn,%
    prl,%
    aps,%
    showpacs,%
    preprintnumbers,%
    amsmath,%
    amssymb,%
    twoside,%
    floatfix,
    reprint, 
    ]{revtex4-1}

\usepackage{graphicx}       
\usepackage{dcolumn}        
\usepackage{color}          
\usepackage{xspace}
\usepackage{tikz}

\newcommand{\ttas}{1T-TaS$_\mathrm{2}$\xspace}
\newcommand{\ttase}{1T-TaSe$_\mathrm{2}$\xspace}

\newcommand{\figref}[1]{Fig.\,\ref{#1}}

\newcommand{\cf}{cf.\xspace}

\newcommand{\ti}[1]{$\mathbf{t}_{#1}$\xspace}

\begin{document}
\title{Disrupted orbital order and the pseudo-gap in layered \ttas}

\author{T.\,Ritschel}
\affiliation{%
Stewart Blusson Quantum Matter Institute,
2355 East Mall, Vancouver B.C., V6T 1Z4, Canada}

\author{H.\,Berger}
\affiliation{Ecole polytechnique Federale de Lausanne, Switzerland}

\author{J.\,Geck}
\affiliation{%
Institute of Solid State and Materials Physics,
TU-Dresden, 01062 Dresden, Germany}

\begin{abstract}
  We present a state-of-the-art density functional theory (DFT) study which
  models crucial features of the partially disordered orbital order stacking in
  the prototypical layered transition metal dichalcogenide \ttas. Our results
  not only
  show that DFT models with realistic assumptions about the orbital order
  perpendicular to the layers yield band structures which agree remarkably well
  with experiments. They also demonstrate that DFT correctly predicts the
  formation of an excitation pseudo-gap which is commonly attributed to
  Mott-Hubbard type electron-electron correlations. These results highlight the
  importance of interlayer interactions in layered transition metal
  dichalcogenides and serve as an intriguing example of how disorder within 
  an electronic crystal can give rise to pseudo-gap features.
\end{abstract}

\maketitle

The spontaneous formation of so-called electronic crystals, i.e., spatial
superlattices formed collectively by the valence electrons of a solid is among  
the most striking topics in 
current condensed matter research~\cite{Monceau2012,Gruner1994a}.  Charge
density wave (CDW) order is a manifestation of such macroscopic collective
quantum states which occurs in a wide range of materials, including doped copper
oxide systems~\cite{Chaix2017,Chang2012}, heavy fermion
systems~\cite{Gruner2017} or layered transition metal dichalcogenides
(TMDs)~\cite{Wilson1975a}. In particular the latter class of materials currently
regains enormous attention as ``post graphene'' quasi two-dimensional
materials. Their unusual electronic properties propel innovative concepts for
applications ranging from miniaturized electronic devices to
quantum computing~\cite{Stojchevska2014,Yoshida2015,Vaskivskyi2016,Ma2016,Pan2017}.  

Owing to the weak bonding between the
layers, the TMDs are traditionally considered to realize rather two-dimensional
electronic systems. However, recent theoretical and experimental work suggests
that, due to orbital order which is intertwined with the CDW, the electronic structure markedly depends on the stacking arrangement of this combined
order in the direction perpendicular to the
layers~\cite{Ge2010,Ge2012,Darancet2014,Ritschel2015,Chen2016,Navarro-Moratalla2016,Hovden2016,Dolui2016}.  
Accordingly, density functional theory (DFT) based on oversimplified
assumptions about the CDW stacking will yield electronic structures that do not agree
with experiment~\cite{Ritschel2015}.
In this letter we present a DFT model that realistically approximates the
experimentally found partially disordered CDW stacking for the prototypical
material \ttas and compare these calculations to angle resolved photoemission
spectroscopy data. \ttas serves here as a prime example to study the effect of
disorder within the orbital sector on the electronic structure.

\begin{figure}[htb]
  \centering
  \includegraphics[width=\columnwidth]{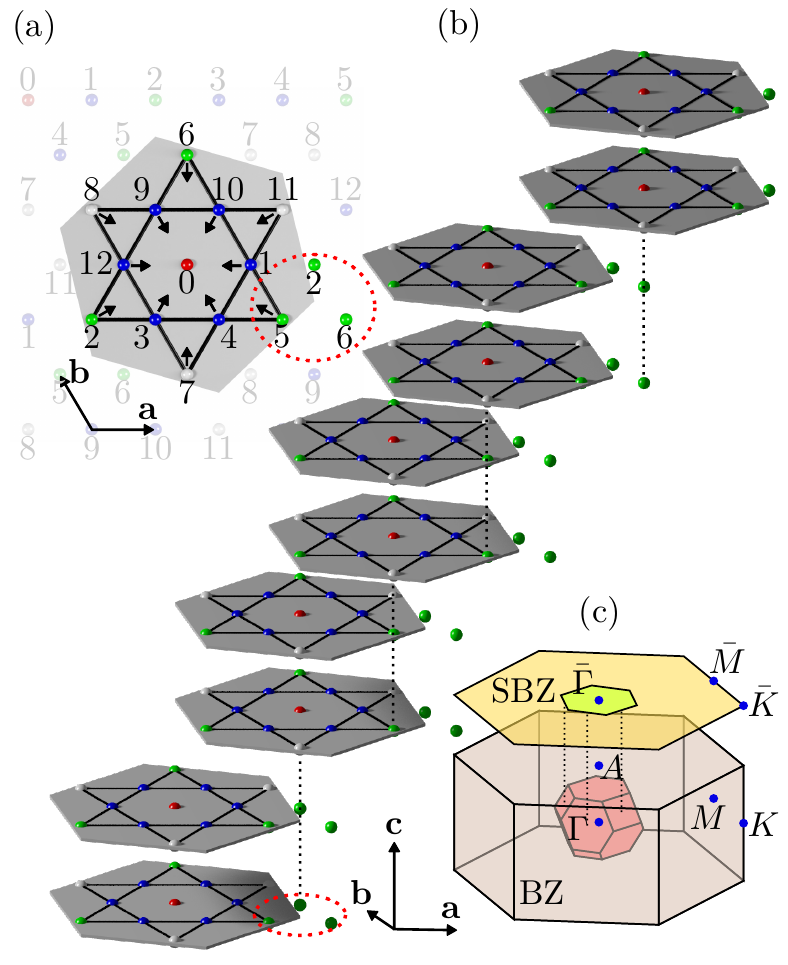}
  \caption{CDW layer stacking of the C-CDW in \ttas. (a): The star-of-david
    shaped in-plane $\sqrt{13}\times\sqrt{13}$ supercell comprises clusters of
    13 Ta sites (gray hexagon). (b): The  \ti{0(2,5,6)} stacking of these
    clusters in the out-of-plane direction is given by on-top stacked bilayers
    which by themselves are stacked by a vector randomly chosen from three
    symmetry equivalent vectors corresponding to the green sites in (a). The
    normal cell lattice vectors are indicate ($\mathbf{a,b,c}$).  (c): The bulk
    Brillouin zone (BZ) and surface Brillouin zone (SBZ) corresponding to the normal cell and the supercell
    of the approximated periodic stacking \ti{02}.}
  \label{stacking}
\end{figure}

Among the TMDs \ttas is well known for its particularly rich electronic phase
diagram as a function of temperature and pressure~\cite{Sipos2008}. Starting at
ambient pressure and low temperatures the in-plane (IP) CDW order is commensurate (C)
with the underlying crystal lattice. It is characterized by star-of-david  shaped
clusters comprising 13 Tantalum sites arranged in a $\sqrt{13} \times \sqrt{13}$
IP superstructure as illustrated in \figref{stacking}~(a). Increasing
the temperature or rising the external pressure yields a so-called nearly
commensurate (NC)-CDW which is associated with the  creation of defects within
the C-CDW which, by themselves, order on an again larger length
scale~\cite{Spijkerman1997}.  At higher temperatures or pressures the
NC-CDW transforms into incommensurate (IC)-CDWs.

Interestingly, not only the IP structure is effected by the transitions
between the different CDWs but also the stacking arrangement in the out-of-plane
(OP) direction. The IC-CDW and the NC-CDW possess a well ordered stacking along
the OP direction with a periodicity corresponding to three times the layer-to-layer distance
$c\approx5.9$\,\AA~\cite{Spijkerman1997}. In contrast, the C-CDW is governed by a complex alternating and partially
disordered stacking along the OP direction: According to the 13 Tantalum sites
forming one star-of-david cluster (labels $0\ldots12$ in \figref{stacking}~(a)), 
there are 13 possibilities of how two
adjacent layers are aligned with respect to each other.
We introduce the following notation:
An arrangement where the central site, i.e. site 0, within a star-of-David
cluster is centered above site $i \in (0\ldots12)$ in the layer below is
referred to as \ti{i}. Taking the threefold IP symmetry of the C-CDW
into account the 13 distinct stacking types form
5 symmetry equivalent groups, namely \ti{(2,5,6)},
\ti{(7,8,11)}, \ti{(1,3,9)}, \ti{(4,10,12)} and \ti0.
Although the stacking of the C-CDW in \ttas is still debated, there is firm
theoretical~\cite{Walker1983,Nakanishi1984} and experimental~\cite{Tanda1984,Ritschel2015}
evidence that it can be viewed as bilayers which are
stacked on top, i.e., stacking \ti0. These bilayers, for their part,
are stacked by a random choice out of the three symmetry equivalent
possibilities corresponding to the group \ti{(2,5,6)}.  The resulting
alternating partially disordered stacking, which we will denote by \ti{0(2,5,6)} is illustrated in
\figref{stacking}~(b). 
To the best of our knowledge, the peculiar disordered C-CDW stacking is only found in
\ttas. The closely related material \ttase develops the same
$\sqrt{13}\times\sqrt{13}$ IP C-CDW structure. However, in this material the
stacking is governed by the formation of macroscopic domains
with a long-range ordered stacking corresponding to one stacking out of the
group \ti{(2,5,6)}~\cite{Wiegers2001}. 
Interestingly, it has been shown, that this
long-range OP stacking order can be transformed into a disordered stacking
among the group \ti{(2,5,6)} by subtle doping of Zr impurities on the Ta
site~\cite{Moncton1976}. However, as opposed to \ttas there is no alternation between
different stacking groups.

Another remarkable feature of the C-CDW in \ttas lies in
its semiconducting transport properties. Commonly, this is  attributed to
Mott-Hubbard type electron-electron correlations. According to this scenario
every star-of-David cluster contributes a single 5d electron to a half filled narrow
conduction band. A sufficiently large Coulomb repulsion $U$ acting on these clusters is
believed to drive a Mott-Hubbard transition with overlapping Hubbard
subbands~\cite{Fazekas1979,Fazekas1980}. 
It has been proposed that Anderson localization finally drives
a metal-to-insulator transition at low temperatures yielding the observed
semiconducting transport properties~\cite{Dardel1992,Dardel1992a}. It should be
noted that a semiconducting CDW phase is very uncommon among the class of TMDs.
All other CDW phases are metallic or even superconducting
at low temperatures~\cite{Rossnagel2011}.
Recently, time and angle resolved photoemission spectroscopy (trARPES) was
employed to observe the ultrafast collapse of an excitation gap at the Brillouin
zone center ($\Gamma$-point, \cf~\figref{stacking}~(c)) -- the so called Mott
gap -- which was interpreted as a clear signature that electronic degrees of
freedom are involved in the formation of this
gap~\cite{Perfetti2008,Hellmann2010a,Hellmann2012}.

\begin{figure*}[tbh]
  \centering
  \includegraphics[width=\textwidth]{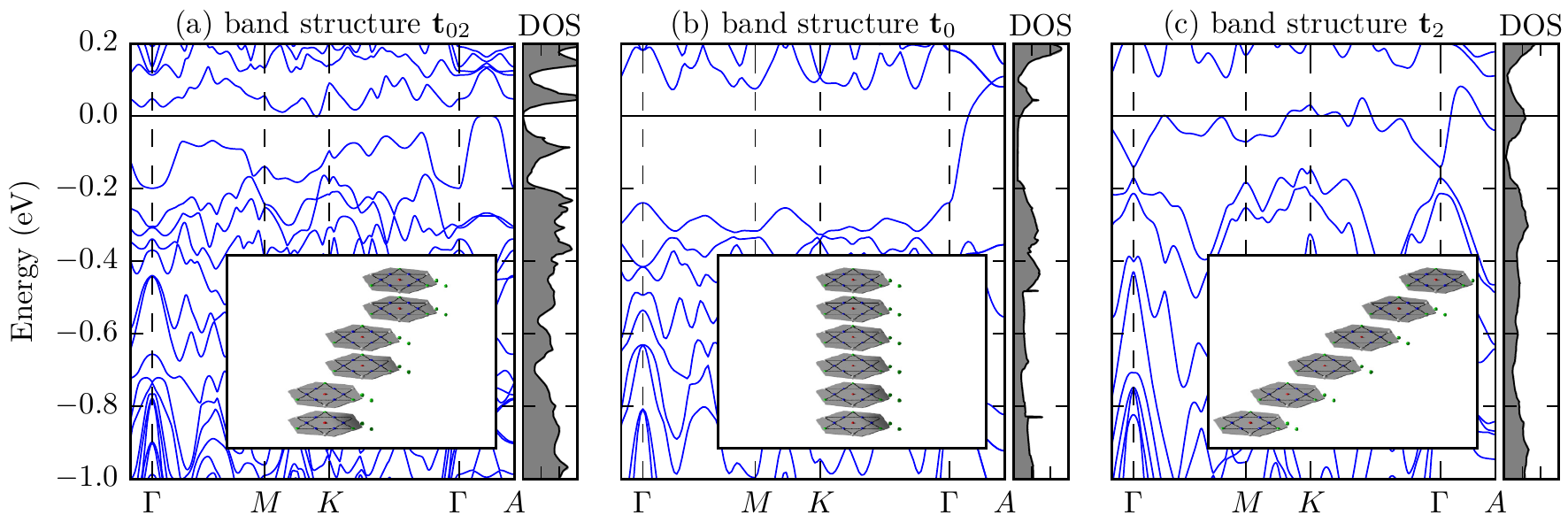}
  \caption{DFT band structure and density of states
  for the same in-plane supercell but different stacking types. (a): The
  alternating stacking \ti{02}. (b): The on-top stacking \ti0. (c): The
  non-alternating stacking \ti2. Illustrations of  the different stacking 
  types are given in the insets, respectively. The high-symmetry points
  correspond to the normal cell Brillouin zone (\cf~\figref{stacking}~(c)).}
  \label{bandstructure}
\end{figure*}

However, in what follows we will use an extension of our previous DFT
approach~\cite{Ritschel2015}
to show that the transport properties and in particular the gap at $\Gamma$ of the
C-CDW phase can be inferred from the actual CDW stacking.
All DFT calculations have been done using the FPLO14 package~\cite{Koepernik1999}.
Following our previous approach the IP supercell structure 
was derived from reference \onlinecite{Brouwer1978} and
\onlinecite{Spijkerman1997}.  In order to incorporate the different stacking types we
constructed triclinic supercells without changing
the atomic displacements and using the experimental interlayer
distance~\cite{Brouwer1978}.
The disordered alternating stacking of the C-CDW as
illustrated in \figref{stacking}~(b) does not obey translational symmetry in
the OP direction and, hence, it cannot be treated in a conventional DFT
bandstructure scheme. In order to circumvent this difficulty
we approximate the disordered stacking 
by a periodic alternation of the stacking \ti0 and \ti2. 
In other words,
the partial disorder among the group of symmetry equivalent stacking types
\ti{(2,5,6)} is neglected and a single fixed stacking 
\ti2 is chosen. We denote this stacking, which is visualized in the inset of
\figref{bandstructure}~(a), by \ti{02}. In
\figref{bandstructure}~(a) we show the result of a DFT bandstructure calculation
for this periodically alternating stacking. For comparison we also reproduce the
bandstructure calculations from Ref.~\onlinecite{Ritschel2015} for the
non-alternating stacking types \ti0 and \ti2 in
\figref{bandstructure}~(b) and (c), respectively.

The first and most important observation is that the DFT calculation within the
local density approximation (LDA) which takes the alternating stacking into account
indeed yields a gap-like feature at the Fermi energy. This
can be most clearly seen from the density of states (DOS) shown in the right
panel of \figref{bandstructure}~(a).  Thus, the key result of these calculations
is that LDA predicts an insulating or semiconducting ground state for the
alternating stacking \ti{02}.  In contrast, the non-alternating stacking types
\ti0 and \ti2  possess a Fermi surface and, hence, our LDA calculations predict
metallic ground states (\cf \figref{bandstructure}~(b) and (c)) in
accordance with previous calculations~\cite{Bovet2003,Ge2010,Darancet2014,Ritschel2015}.

Although the formation of this gap is surprising at first glance it can be
understood in terms of elementary band theory arguments: The supercell corresponding to
stacking \ti0 or \ti2 contains an odd number of electrons (13 Tantalum sites
each contributing one 5d electron). LDA calculations for these
stacking types result in metallic ground states as the valence band is
partially filled. In order to model the alternating stacking it is
necessary to include two star-of-David clusters in the supercell.
Accordingly, the resulting even number 
of electrons in this supercell can yield a completely filled valence band and
thus an insulating or semiconducting ground state. However, this line of arguments does
certainly not allow to deduce the
size of the gap. Depending on subtle details of the hopping integrals between
wavefunctions in adjacent layers the gap size could indeed be zero
or there could even be a small overlap of valance and conduction band resulting
in  a semimetal with conducting behavior at finite temperatures. 
Interestingly, the band structure and DOS shown in \figref{bandstructure}~(a) suggest
exactly such a zero-gap semiconductor for the case of the
\ti{02}-stacking. Measurements of 
the Hall effect in \ttas find semiconducting or semimetallic behavior in the
C-CDW phase of \ttas, which supports our theoretical result~\cite{Inada1980}.

\begin{figure*}[htb]
    \centering
    \includegraphics[width=1.\linewidth]{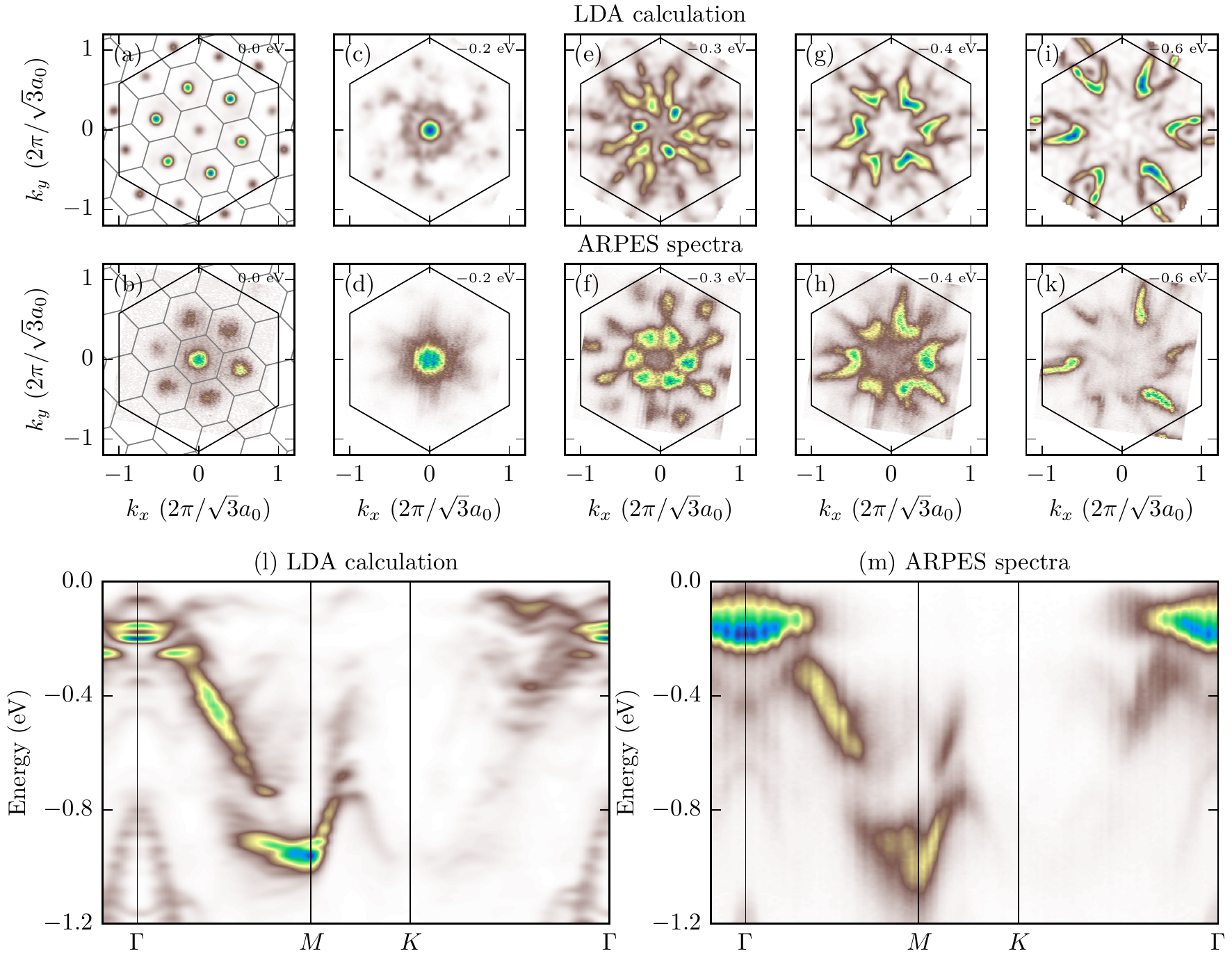}
    \caption{Comparison of the simulated spectral function corresponding to the
        \ti{02} stacking to ARPES data. (a) - (k): Constant energy cuts
        (energy is indicated in the top right corner of the graphs).  Thick and
        thin solid lines in (a) and (b) indicate the surface Brillouin zone
        boundaries corresponding to the normal cell and supercell, respectively.
        $a_0 = 3.36$\,\AA{} is the in-plane lattice parameter of the undistorted
        crystal structure. (l) and (m): Cuts along high symmetry directions of the
        normal cell surface Brillouin zone (\cf~\figref{stacking}~(c)).}
    \label{ARPES_DFT}
\end{figure*}

We will now assess the validity of our LDA model of the C-CDW phase by a 
detailed comparison of the 
calculated electronic band structure 
to angle-resolved photoemission spectroscopy (ARPES) measurements. To this end
we have measured the valance electronic structure of a high quality single
crystal of \ttas as a function of $k_x$ and $k_y$ which, together with the
energy-axis, yields a 3D data set. 
These measurements were performed at the $1^3$-ARPES endstation at beamline
UE112PG2 of the Berlin Synchrotron (BESSY).  We used $p$-polarized light of
96\,eV photon energy, so that the final state crystal momentum at normal
emission corresponds to the $\Gamma$-point\,\cite{Rossnagel2005}.
The sample temperature was kept at 1\,K.
In \figref{ARPES_DFT} we compare various
cuts through the obtained 3D data set 
with corresponding DFT calculations.
We use the unfolding scheme which is described in Ref.~\onlinecite{Ku2010}. This approach
facilitates a comparison of the theoretical bandstructures to the experimentally accessible spectral
function $\mathcal{A}(k,\omega)$~\cite{Damascelli2003}. Furthermore we use a heuristic approximation
in order to mimic the effect of the partial disorder which is a marked
characteristic of the CDW 
stacking in \ttas: Since the three vectors \ti2, \ti5 and
\ti6 occur
randomly in the partially disordered stacking the whole structure obeys,
on average, threefold rotational symmetry. This symmetry is broken by the
approximation \ti{02}, where only
\ti2 occurs.  For a comparison to ARPES data we have therefore
symmetrized the calculated unfolded bandstructure for \ti{02}
in order to restore the threefold symmetry.
This symmetrization  bears on the assumption  that, on average, the effect of
the disorder on the electronic structure can be approximated with
a linear incoherent superposition of \ti{02}, \ti{05} and \ti{06}.
In addition, it is known that ARPES spectra often represent an inherent
$k_z$ integration of the electronic structure~\cite{Koitzsch2009}. In other
words, the $k_z$ momentum of the electrons  which are probed by an APRES
experiment is not sharply defined. Instead also electrons with different $k_z$
momenta contribute to the photocurrent.  This might be even more relevant for
partially disordered structures since, strictly speaking, $k_z$ does not
represent a valid quantum number in such a system.  In order to simulate this
effect
we consider a weighted linear combination of unfolded IP bandstructures
corresponding to six equidistant $k_z$ values between $\Gamma$ and the midpoint between $\Gamma$ and $A$. The
weighting factors follow a Gaussian distribution centered at $\Gamma$.
Finally, the unfolded bandstructure is convoluted with a resolution function in
order to obtain the simulated spectral function shown in \figref{ARPES_DFT}. Note that the present LDA-simulation does not account for more elaborated matrix
element effects~\cite{Lindroos2002} of the photoemission process, causing deviations between  LDA-simulation and experiment. However, these deviations play no role for the following discussion.

In fact, apart from details in the intensity distribution, the overall agreement between measurement and calculation in \figref{ARPES_DFT} is remarkable.
For instance, not only the energy of the electron pocket feature at the 
$\Gamma$-point is well reproduced by our calculations (\figref{ARPES_DFT}~(l,m)).  
Also the shape of that feature is nicely described by the LDA model. This
is remarkable as these bands are rendered hole-like for the non-alternating
stacking types (\figref{bandstructure}~(b,c)). 
It should be emphasized that the gap between the center of this feature at the $\Gamma$-point and the Fermi level is
commonly explained in terms of electron-electron correlations and is therefore often referred to as the ``Mott-gap''. This interpretation is indeed corroborated by the ultrafast response of this gap to photo-excitation, which appears inconsistent with a gap caused by electron-phonon interactions~\cite{Perfetti2008,Hellmann2010a,Hellmann2012}.

Notwithstanding, the quantitatively correct prediction of this feature by LDA strongly argues against electron-electron correlations causing the gap at the $\Gamma$-point. Instead, our results indicate that this gap is dominated by interlayer  
hybridization. The marked sensitivity of the low-energy electronic structure with respect to the OP order found here, is indeed rooted in orbital textures, which are
interwoven with the CDW~\cite{Ge2010,Ritschel2015}. Different stacking
arrangements alter the hopping integrals between the orbitally ordered layers in
a non-trivial manner yielding pronounced changes in the electronic structure. It is important to point out that the electronic orbitals can respond to external perturbations on electronic time scales, i.e.\ our result is in keeping with the observed ultra-fast response of the gap at $\Gamma$.

The energy cuts at the Fermi level (Fermi surface) shown in
\figref{ARPES_DFT}~(a,b) bear another interesting result: 
The weak signal observed in ARPES was previously referred to as a remnant or pseudo-gapped Fermi surface~\cite{Dardel1992,Pillo1999,Bovet2004,Borisenko2008}. 
Surprisingly, similar features occur in our LDA calculations (cf. \figref{ARPES_DFT}~(a)). Note that the finite intensity at the $\Gamma$-point in the calculation is a result of the $k_z$ integration: The electron pocket at the $\Gamma$-point transforms
into an hole-like band as $k_z$ increases towards the $A$-point. This hole like
band can be seen in \figref{ARPES_DFT}~(l) and (a) since the simulated $\mathcal{A}(k,\omega)$ also contains these bands as a result of the $k_z$ integration.
LDA calculations for the alternating \ti{02} stacking naturally produce bandstructures corresponding to a zero-gap semiconductor (\cf \figref{bandstructure}~(a)). The pseudogap feature, which is observed in ARPES at photon energies nominally
corresponding to the $\Gamma$-point, might therefore  be directly related to the
partial disorder of the stacking in the real material, as this breaks the translational symmetry in the OP-direction, causing a pronounced broadening in the $k_z$-direction and yielding similar effects as the $k_z$-integration performed here.

In conclusion, we presented LDA-calculations for the C-CDW phase of \ttas, which  reproduce all the main features of the experimentally observed electronic structure on a quantitative level. They key ingredient for the presented supercell calculations is a realistic description of the OP stacking.
Strong electron-electron correlations are therefore not necessary to understand the electronic structure of \ttas in the C-CDW phase.
In particular the so-called Mott gap appears not to be driven by strong electron-electron interactions. Instead this gap can be explained \textit{quantitatively} by the hybridization of orbitally ordered $ab$-planes stacked along the $c$-direction. As a result, despite the layered structure of the TMDs, the OP stacking plays a dominant role for the IP electronic (gap) structure and, hence, the transport properties. At the core of this strong interplay between electronic band structure and CDW-stacking lies the complex orbital order, 
which was previously identified to form along with the CDW in TMDs~\cite{Ritschel2015}. 
Finally our results provide an example of how disorder within an electronic crystal (disordered stacking of orbital order) can give rise to pseudo-gap features -- an effect that may also be relevant to the much discussed pseudo-gap phase of the high-temperature superconducting cuprates. Such orderings of orbital degrees of freedom and related disorder effects might therefore impact a wide range of materials and certainly deserve further scrutiny in future studies.

This work was financially supported by the Deutsche Forschungsgemeinschaft under
Grant No. RI 2908/1-1, DFG-GRK1621 and GE 1647/2-1. We thank K. Rossnagel and G.
Sawatzky for fruitful discussions.

\end{document}